\documentclass[twocolumn,prd,aps,amssymb,amsmath]{revtex4}
\usepackage{graphics}
\usepackage[pdftex]{graphicx}
\newcommand{\ket}{\rangle}
\newcommand{\bra}{\langle}

\begin{document}
\preprint{}

\title{
Minimum Decision Cost for Operators
}
%and a description of implementation proposals}

\author{Bernhard K. Meister}

%\affiliation{...., Imperial College, London SW7
%2BZ, UK}
\affiliation{ Department of Physics, Renmin University of China, Beijing, China 100872}
\email{ bernhard@fasteagle.jp}
\date{\today }

\begin{abstract}
%The binary decision problem for operators with a given prior is investigated. Results developed for the well studied problem of non-orthogonal state discrimination with the aim of minimizing the error probability are utilized.

%Bounds are provided for the error probability for all binary operator choices, and it is shown how probe entanglement enhances the result.
 
 State discrimination with the aim to minimize the error probability is a well studied problem. Instead, here the binary decision problem for operators with a given prior is investigated. A black box containing the unknown operator is probed by selected wave functions. The output  is analyzed with conventional methods developed for state discrimination. An error probability bound for all binary operator choices is provided, and it is shown how probe entanglement enhances the result.

%The aim is to optimally  distinguish between different possible operators  enclosed in a black box tested by selected input wave functions.
%In this paper the Bayes version of the binary decision problem is studied and it is shown how entanglement can used to %reduce the error probability. The set up is the following, an input probe is sent through a black box, which either %leaves the probe unchanged or modifies it in a fixed way, e.g. changes the polarization of the probe. The resulting %state is analyzed by conventional optimal state discrimination
% methods based on the Bayes cost. The main result of paper is to show how entanglement of the input probes reduces the %minimal cost.

\end{abstract}
%\pacs{PACS Numbers : 02.50.-r, 02.50.Le, 03.65.Bz}
\maketitle

%{\it Keywords:}  Quantum measurements;

\section{Introduction}
%\label{sec:1a}
Quantum metrology encompasses quantum decision problems and has gained prominence with the rise of quantum engineering. One common task is to determine an optimal observational strategy. This has been carried out in various frameworks. Here we focus on the Bayes procedure, which is natural if prior information is available.
This leads one to seek a strategy minimizing  the expected cost, or in other words to minimize the error probability. The cost function is chosen to be $0$ - $1$, where $0$ is associated with the correct and $1$ to the incorrect choice.
 A detailed exposition of the general quantum decision strategy for wave functions can be found in
the book of Helstrom %\cite{hel}
[1], papers by Holevo [2]
%\cite{holevo} 
and Yuen {\it et al.} [3]  
%\cite{yuen}
,  where independently much of the groundwork of the field was laid.
%As the next paragraph describes one cannot only apply decision analysis to states but also to operators.

%Non-orthogonal state discrimination  with the aim to minimize the error probability is a well studied problem. Here instead the binary decision problem for operators with a given prior is examined. The aim is to optimally  distinguish between different possible operators  enclosed in a black box acting on selected input wave functions. The output of the black box is analyzed with conventional methods developed for state discrimination. Bounds are provided for the error probability for all binary operator choices, and it is shown how probe entanglement enhances the result.

In the present paper instead of distinguishing states  we  optimally distinguish between operators.  We study the Bayesian approach to a binary decision problem for a probe made up of an ensemble of particles being sent through a black box containing an operator selected from a set of two with known prior. The insight employed is that the binary operator decision problem can  be mapped into a binary state decision problem. This mapping is allowed, because the unknown operator in the black box transforms the input state into one of two possible output states.  The two possible output states have a well-defined transition probability, which is all one needs to apply the standard state decision machinery. The challenge is to find the optimal state in the set of all allowed  inputs to obtain the smallest transition probability in the output states. To solve this we use techniques previously applied to the quantum brachistochrone. %This set up is slightly different from the  earlier paper \cite{brody}, where the states to be distinguished either as a ensemble of N identical prepared particles were directly provided to the experimenter, who could either measure them sequentially or together.

The  operator decision problem is interesting in its own right, since the space of operators has a Finsler metric and is the natural space for quantum algorithms, e.g. in Grover's algorithm different positions of the marked state are associated with different operators acting on an initial state, and one can rewrite the algorithm as a multiple operator decision problem[4].%\cite{bkm}.

%What was shown rather surprisingly in the earlier paper is that in the quantum case the optimal minimal cost is the same for sequential or one big joint measurement.
%Here instead we try to distinguish not input states but operators in 'black boxes'.

A relevant application of the Bayesian approach to quantum hypothesis testing for multiple polarised spin 1/2 particles was given in Brody {\it et al.} [5].%\cite{brody}. %As part of the quantum decision strategy any positive operator valued %measure (POVM) was allowed.
Two strategies were sketched out and compared. In one the experimenter is provided the input as a sequence of identical particles with an adjustable measurement for each particle and in the other one joint measurement is carried out on all particles.
%The Bayes sequential analysis to the ensemble of particles.
%The result is then compared with a combined measurement of the entire ensemble, treated
%as a single composite system.
It was shown that
the optimal Bayes cost for separate sequential measurements of the individual particles is the same
as that of a combined measurement. Intermediate strategies were also shown to be unable to reduce the Bayes cost. 
%This could be seen as surprising, since
%%the sequential allows for adjustments in the measurement direction incorporating the earlier outcomes, but
%being able to adjust the measurement apparatus after each particle as is possible in the sequential case provides no benefit. 
% Other strategies consisting of combined measurements of
%sub-ensembles are also considered.
%Any other strategy turns out to entail a higher expected cost.

In the next section the Bayes cost calculation will be carried out for two of the simplest possible operators acting on an input, i.e. either the $2$-dimensional identity operator or an operator which leaves one of the two eigenstates unchanged and adds a phase shift to the other eigenstate. It is shown how entanglement reduces the Bayes cost. In the penultimate section the binary decision problem is extended to a larger class of operators. In the last section the results will be summarized and some additional points briefly discussed.

\section{Example Calculation in the Binary Operator Decision Case}

In this section one particularly simple binary operator decision problem, generalized in the next section, is described to clarify the procedure. The calculation is divided into three subsections. One each dealing with  the entangled and unentangled case, and in the last subsection the results are compared.

As part of the set up we choose a
 diagonal 2-dimensional operator  of the form
 \begin{eqnarray}
U_i=\left(
\begin{array}{cc}
1 & 0\\
0 & e^{\imath \alpha_i} \\
%\vdots & \vdots & \ddots & \vdots\\
%0 & 0 & \cdots & E_n\
\end{array}\right), \nonumber%
%\right)\neq
\end{eqnarray}
with $\alpha_1=0$ and $\alpha_2=2\delta$.
% corresponding the having the Hamiltonian
%\begin{eqnarray}
%H_i=\left(
%\begin{array}{cc}
%0 & 0\\
%0 & \alpha_i \\
%\end{array}\right), \nonumber
%\end{eqnarray}
%applied (corresponding to a phase shift gate made up of an overall phase shift and $sigma_z$ in the $i=0$ case and
%and just an overall phase (which is ignorable) in the other case .
%This example  is useful for understanding the method applied and leads to a generalization developed in the next section. 
The prior probability is given by $\xi$ for the first    and hence $1-\xi$ for the other case. The experimenter has to find out by sending probes through the black box, if it contains $U_1$ or $U_2$.
 The advantage of this set up is that a closely related  binary state decision problem has been analyzed  in Brody {\it et al.} % \cite{brody}
 [5], and the remaining challenge is to understand how entanglement effects the transition probability.

The decision strategy has various parts.
 %now to be discussed in turn. 
 One has to decide on the probe state, on the
multiple interactions between parts of the probe  and the black box, and on the multiple measurements of the output to optimally determine the content of the black box. The whole range of positive operator valued measures are allowed in the decision strategy as was the case in the earlier paper. %The procedure is described in the next paragraph.
 %The  black box considered either leaves the probe unchanged or modifies it by changing the phase of one of the operator eigenfunctions in a prescribed way. 
 
 Next more details are provided about the components.
 The experimenter  can choose as a probe a wave function freely from the set of $N$ particle ensembles. These $N$ particles can be entangled or sent as a product state. Naturally there are some constraints on the allowed input states, e.g. the dimension of the Hilbert space of the input particles and the black box Hamiltonian have to match. In the current example each particle is represented by a $2$-dimensional wave function.
 In addition to the  probe the experimenter is provided with one black box that can be used multiple times, but at most $N$-times, to modify the probe. As an additional constraint each particle, entangled or not, can at most be sent through the black box once\footnote{A single particle state can reach the $N$-particle entangled state error probability limit, if it can pass through the black box $N$-times.}.
 %The black box output can  tested sequentially, with the ability to use the earlier measurement outcomes to modify later measurement arrangements, or measured together.
%   The aim of the experimenter is to ascertain, what kind of black box the probes were sent through. 
 The final step is to choose how the probe after passing through the black box is measured and how the result is analyzed, i.e. in terms of preposterior analysis.
 
 One should keep in mind that in principle part of probe preparation as well as probe and black box interaction can be influenced by earlier measurements. This would complicate the situation, since partial measurements with binary outputs followed by passing the probe again through the black box increases the number of possible outputs.  Here instead we assume that probe and black box interaction are completed before any measurement takes place.
 %One should also keep in mind the resource constraints, which are the number of particles contained in the probe and the number of times a particle, which is part of the probe, is sent through the black box. In the subsequent section we always use $N$ to represent the number of probe particles.
% In the next section the same problem is analyzed for an arbitrary pair of possible operators.

%REDUNDANT:

%Let us talk next about the allowed probes before we discuss the black box further.
%The input is a combination of 2-dimension particles.  In the one particle case, e.g. a spin 1/2 particle of arbitrary polarization. In the multi-particle case one can now prepare them either as a simple product state, as fully entangled, or as combination of entangled and product states.
%After the state preparation comes the interaction between probe and black box. Next one passes each particle, once   through the black box.

%All these steps are carried in the next subsection for the initially defined $U_i$ operator.  "

%The testing procedure is divided into the state preparation part, the state modification part, and
%two parts. %First one has to choose the set of allowed probe.
%First the input states are chosen, then they are sent through the black box either sequentially interspersed with measurements or jointly.

%. Finally the
%In our case the experimenter first has to choose how many particles are to used.

\subsection{Probe: One or more unentangled particle}

We consider first the simplest case of sending just {\it one} particle through the black box. Any
2-dimensional wave function of the form  $(a|0\ket + b|1\ket)$ with $|a|^2+|b|^2=1$ is allowed as an input. %and it can be shown that each should have equal weight.
The two possible outputs are $(a|0\ket + be^{i \alpha_i}|1\ket)$.
The transition probability between the possible outputs is
\begin{eqnarray}
&&(|a|^2+|b|^2e^{i 2\delta})(|a|^2+|b|^2e^{-i 2\delta})\nonumber\\
%&=&|a|^4+|b|^4+2 |a|^2|b|^2   \cos(2\delta) \nonumber\\
%&=& |a|^4+|b|^4+2 |a|^2|b|^2 (2 \cos^2(\delta)-1)\nonumber\\
&=&1 - 4 |a|^2|b|^2\sin^2(\delta),\nonumber
\end{eqnarray}
which is minimal for $|a|^2=|b|^2=1/2$
leading to a transition probability of $\cos^2(\delta)$.

After the one particle probe has been successful sent through black box, and either changed or left unchanged, the next step is to carry out a measurement. An operator decision problem has been turned into a problem of distinguishing between two different wave functions.

 The optimal procedure for states was developed by Helstrom and others, and  the optimal cost function, also called the Helstrom bound, has  for a transition probability of $\cos^{2}( \delta)$ and prior $\xi$ the form
\begin{eqnarray}
C_{unEnt}(\xi,1)= \frac{1}{2} \Big(1-\sqrt{1-4 \xi(1-\xi) \cos^{2} (\delta)}\Big).\nonumber
\end{eqnarray}
 The strategy  minimizing the error probability between the output states also minimizes the error in the operator decision problem.

Next we consider the multi-particle case.
 Two types of probes are considered.  First, a sequence of $N$ independent 2-dimensional particles each of the form $1/\sqrt{2}(|0\ket + |1\ket)$;
second we take the direct product state $(1/\sqrt{2}|0\ket +1/\sqrt{2} |1\ket)^{\bigotimes^N}$.
As a result the total transition probability is in each case $\cos^{2N} (\delta)$, i.e. either one has $N$-times a transition probability of $\cos^{2} (\delta)$ or in the product state case one transition probability of $\cos^{2N} (\delta)$.

%If the experimentor has only one particle to ...
%Now in the multiple particle case we can either send in a sequence of $N$ individual particles or the direct product state of the form
%$1/\sqrt{2^N}(|0\ket + e^{\i  \alpha_i}|1\ket)^N$.

Following very closely the argument in a paper by Brody {\it et al.} [5] %\cite{brody}  % and augment it by allowing the use of entangled states
one can carry out either a sequence or a combined measurement to optimally determine between the 2 cases.
This leads in both cases to
%The original results by Helstrom showed that the
an optimal binary decision cost of
\begin{eqnarray}
C_{unEnt}(\xi,N)= \frac{1}{2} \Big(1-\sqrt{1-4 \xi(1-\xi) \cos^{2N} (\delta)}\Big).\nonumber
\end{eqnarray}
This finishes the analysis of the unentangled case.  In the next subsection we deal with entangled probes.

\subsection{Probe: Entangled particles}
The question, if entanglement, a term introduced by Schr{\"o}dinger as `Verschr{\"a}nkung' in the thirties, can be utilized to reduce the minimal decision cost, is of interest due the ongoing fascination with the concept, which now lies at the core of the burgeoning field of quantum information theory.
Therefore, we next consider the cost of the fully entangled $N$ particle state.% $1/\sqrt{2}(|0\ket^{\bigotimes^N} + |1\ket^{\bigotimes^N})$.

Following very closely the argument for product states one can show that
the optimal entangled N particle probe state to maximize the transition probability between the possible outcomes is $1/\sqrt{2}(|0\ket^N + |1\ket^N)$.
The resulting output states are therefore %$1/\sqrt{2}(|0\ket^N + e^{\i N \alpha_i}|1\ket^N)$
\begin{eqnarray}
U^{\bigotimes N}_i \frac{1}{\sqrt{2}}\Big(|0\ket^{N} + |1\ket^{N}\Big) =\frac{1}{\sqrt{2}}\Big( |0\ket^N + e^{\imath N \alpha_i} |1\ket^N\Big)\nonumber
\end{eqnarray}
with $i$ taking the value one or two.
As a result the transition probability between the possible two outputs is $\cos^{2} (N\delta)$
and the associated decision cost is
\begin{eqnarray}
C_{Ent}(\xi,N)= \frac{1}{2} \Big(1-\sqrt{1-4 \xi(1-\xi) \cos^{2} (N\delta)}\Big).\nonumber
\end{eqnarray}
%In principle there could be a partial measurement strategy for the entangled states followed by further modifications to reduce the cost. %but it is not clear how this could be done. 
If probe modification is completed before any  measurement %of the POVM type 
takes places, then the bound applies without restriction.
%It is clear for example that the product state constructed of two  independent entangled states has higher cost than the larger entangled state, since for the two transition probabilities $\cos^2(\delta_1)$ and $\cos^2(\delta_2)$ associated with the product of the two entangled states
% the joint cost is $\frac{1}{2} \Big(1-\sqrt{1-4 \xi(1-\xi) \cos^{2}(\delta_1)\cos^{2}(\delta_2)}\Big)$.
A comparison of the two cost functions follows next.
%or $\cos^{2} (N \delta)$
%Since we for small $\delta$ and  $N \delta $
%the expansion

%\begin{eqnarray}
%U^{\bigotimes N} (|0\ket^{N} + |1\ket^{N}) = |0\ket^N + e^{\imath N \alpha_i} |1\ket^N\nonumber
%\end{eqnarray}
%The choice is now to

%The resulting states is therefore
%$1/\sqrt{2}(|0\ket^N + e^{\imath N \alpha_i}|1\ket^N)$
\subsection{Comparison of the unentangled  and entangled case}
%or an entangled state $1/\sqrt{2}(|0\ket^N + |1\ket^N)$.
%In this subsection we compare the cost in the unentangled and entangled case.
If one compares the two cost functions $C_{UnEnt}(\xi,N)$ and $C_{Ent}(\xi,N)$, one notices that they only differ in the size of the transition probability, which is either $\cos^{2N} (\delta)$ or $\cos^{2} (N \delta)$.
In the case of $N \delta\ll \pi/2$ we have $\cos^{2N} (\delta)\geq\cos^{2} (N\delta)$ and as a consequence
$C_{UnEnt}(\xi,N)\geq C_{Ent}(\xi,N)$. Therefore the optimal entangled cost is always lower or equal than what is possible for any product state with the same number of particles, if the candidate operators are `close' together, i.e. the operators are difficult to distinguish and the black box output transition probability for any input is  small.

After having dealt with the two extreme cases of either a fully entangled or unentangled states (either in sequence or as a product state), one can study if any intermediate case can give a better outcome. This cannot be the case for unentangled states as shown in Brody {\it et al.} [5]. %\cite{brody}. 
In the entangled case for the same reason this is also impossible. %A short calculation can show that no immediate strategy,  entangling  only some of the particles,  can lead to any improvement.
This follows from the following two facts. The transition probability of product   states is the product of the transition probability, and second that for appropriately small product of $\delta$ and $N$, i.e.  for $N\delta\ll \pi/2$,  the following always holds
\begin{eqnarray}
\prod_{j=1}^{K}\cos^{2n_j}(m_j\delta)\geq \cos^{2}(N\delta)\nonumber
\end{eqnarray}
for $N=\sum_{j=1}^{K}n_j m_j$.
This immediately means that standard partial measurement strategies, even optimally implemented, always increase the cost,
since
\begin{eqnarray}
C_{Ent}(\xi,N)\leq \frac{1}{2} \Big(1-\sqrt{1-4 \xi(1-\xi)\prod_{j=1}^{K}\cos^{2n_j}(m_j\delta) }\Big).\nonumber
\end{eqnarray}
This completes the calculation for this simple example. Subsequently, we build on this result.
%The choice is now to

\section{Analysis of the General Binary Operator Decision Problem}

In this section we extend the earlier results to the general binary operator decision problem, where the decision is between any two operators of the same dimension.
%What can be done for the just discussed specialized
 %diagonal 2-dimensional operator
 %of the form
 %\begin{eqnarray}
%U_i=\left(
%\begin{array}{cc}
%1 & 0\\
%0 & e^{\imath \alpha_i} \\
%%\vdots & \vdots & \ddots & \vdots\\
%%0 & 0 & \cdots & E_n\
%\end{array}\right), \nonumber%
%%\right)\neq
%\end{eqnarray}
%can be extended to any decision problem between two operators of the same dimension,
The two operators are again called $U_1$ and $U_2$ and the prior is still $\xi$ and $1-\xi$. The aim is to find a state $\phi$ for which the transition probability between the two possible output states
\begin{eqnarray}
|\bra \phi | U_0^\dag U_1 |\phi\ket|^2
\end{eqnarray}
is minimal. The reason, why this is sufficient to solve the decision problem is the same as in the special example discussed above. % and where $U_0$ and $U_1$ are the two possible operators.
%To be able to solve this problem it is necessary to study the
%structure of $U_1^\dag U_2$. %with particular attention to the eigenvalues. %, since the difference between the largest and smallest Eigenvalue puts a bound on the
%speed limit of the evolution.
Instead of tackling the problem directly, let us rewrite the operator   $U_i$ as $e^{\imath H_i t/\hbar}$, where $H_i$ is the related Hamiltonian, i.e. a Hermitian operator, and $t$ is a real parameter.  Here we choose $t$ to reflect the time a one particle probe spends in the black box. For the rest of the paper $\hbar$ is set to one.

Next we exploit the similarity between our problem and the brachistochrone.
 Here we really study  the dual of the brachistochrone,  where instead  of the optimal Hamiltonian within the constraints, one has to find the optimal input state within the constraints.  In the brachistochrone problem eigenvalues play an important role. In particular it is true for all Hamiltonians that an equal weighted superposition of the eigenstates with largest and smallest eigenvalue evolves faster away from the input state than any other state, i.e. the resulting output state has the smallest transition probability to the input state for small time intervals.  For a concise geometric derivation of the speed limit\footnote{In generalized PT-symmetric quantum mechanics this speed limit does not hold, see Bender {\it et al.} [7].} %\cite{bender}
%governed by the difference between the maximal and minimal eigenvalue of the Hamiltonian
in  the brachistochrone  problem and  the connection to the Anandan-Ahrononv relation see Brody [6]. %\cite{brody2}
 In  our setting this means that an equal superposition of any maximal eigenvalue and minimal eigenvalue eigenstate is the optimal input state for minimizing the output transition probability.

In general, one can for any unitary $U_i$  write $U_1^\dag U_2=U_{new}=e^{\imath H_{new} t}$ with   $H_{new}$ Hermitian and  $t$ the time spent inside the black box.
The largest and smallest eigenvalue of $H_{new}$ are called   $\lambda_{max}$ and $\lambda_{min}$ respectively. Therefore, the optimal state in the one particle case is an equal superposition of the eigenstate with largest and eigenstate with the smallest eigenvalue. The optimal state can be highly non-unique, if there is degeneracy in the eigenstates.
%(and in the case of non-maximal rank one of the eigenvalues can be set to zero, if this increases the difference and therefore decreases the bound).
Independent of the size of the Hilbert space the problem reduces, once the minimal and maximal eigenvalue state has been chosen, to two dimensions. The transition probability between the two possible output states is  $\cos^2((\lambda_{max}t-\lambda_{min}t)/2)$. %, if the rank of $H_{new}$ is maximal.
%In the case where the rank of $H_{new}$ is not maximal, than  either $\lambda_{min}$ or $\lambda_{max}$ can be set to zero, if the spectrum is all positive or all negative respectively.
 To clarify the result, one can just look at the special example analyzed initially, where the black box either does or does not produce a phase-shift. In that particular case the matrix $H_{new}$ was of rank one, $t$ was set to $1$, $\lambda_{max}$ was $2 \delta $, and  $\lambda_{min}$ was zero.

Next, we look at the special case where the parameter $t$ is  small, then
\begin{eqnarray}
|\bra \phi | U_1^\dag U_2 |\phi\ket|^2
\end{eqnarray}
simplifies to
\begin{eqnarray}
|\bra \phi | e^{-\imath H_1 t+\imath H_2 t+\frac{1}{2}[H_2,H_1]t^2+O(t^3)} |\phi\ket|^2
\end{eqnarray}
according to the Baker-Campbell-Hausdorff formula.
For the minimal transition probability one has to find the lowest and highest eigenvalues of the matrix %$\mathbb{I}
$H_{new}=H_1 t  - H_2 t+O(t^2)$, which can again be viewed as a Hamiltonian in its own right. %with its own largest and smallest eigenvalue. %or more simply $[H_0 ,H_1] t^2+O(t^3)$.
%since the identity matrix does not change the direction of the eigenvectors and the shift in eigenvalues is irrelevant %due to sole dependence on eigenvalue differences.
%while ignoring higher order terms in $t$.
If the expansion of $H_{new}$  up to linear order in $t$ is not at least rank $1$, then one needs  to  further expand the matrix by including  higher order terms like $\frac{1}{2}[H_2,H_1]t^2$, and more if necessary, until one obtains at least a  rank $1$ matrix.

Once the optimal minimal transition probability is known in the one particle case, one can, in an analog way to what was done in the first example, extend the result to multi-particle product and entangled states.
%, one can follow the procedure described above for the special example.
The benefits associated with entanglement again become evident. The general minimal cost function is for unentangled states
\begin{eqnarray}
&&C_{UnEnt}(\xi,N)=\nonumber \\
&& \frac{1}{2} -\frac{1}{2}\sqrt{1-4 \xi(1-\xi) \cos^{2N}((\lambda_{max}-\lambda_{min})t/2) },\nonumber
\end{eqnarray}
and if entangled states are allowed it becomes
\begin{eqnarray}
&&C_{Ent}(\xi,N)=\nonumber \\
&&\frac{1}{2} -\frac{1}{2}\sqrt{1-4 \xi(1-\xi) \cos^{2}(N (\lambda_{max}-\lambda_{min})t/2)}.\nonumber
\end{eqnarray}
%where as stated above either $\lambda_{max}$ or $\lambda_{min}$ can be set to zero, if $H_{new}$ is not maximal rank.
%In the non-maximal rank case one sets the eigenvalue to zero that has the smaller absolute value.
As before the entangled cost is smaller than the entangled cost as long as $N(\lambda_{max}-\lambda_{min}) t\ll \pi$, i.e. $C_{Ent}(\xi,N)\leq C_{UnEnt}(\xi,N)$. 
This concludes the analysis of the general case, even if one can with the help of the well-developed machinery of matrix theory describe interesting properties of the minimal and maximal eigenvalues of $H_{new}$, and analyze special cases, where $H_{new}$ is of non-maximal rank. This will be done in a different setting, where the focus will be on  applications.

%As an example choose $H_0$ to be the Pauli matrix $\hbar\sigma_x/2$ and $H_1$ to be the Pauli matrix $\hbar\sigma_y/2$ then the
%commutator is $ i\hbar \sigma_z$ and the eigenvalue difference $2\hbar t \sqrt{2 - t^2}+O(t^3)$ and the transition probability is $\cos^2(\hbar t\sqrt{2})$ up to order $O(t^3)$ between the two possible probe output states.
%and the other extreme

%where
%\section{Part IIa:  Optimal measurements for achieving lowest cost in the fault-free state separation case}

\section{Conclusion}
The paper solves the binary operator decision problem  by translating it into the well understood problem of distinguishing states. Once this is recognized, the rest of the analysis is mainly cranking the handle of a well-oiled machine.

The general problem of distinguishing arbitrary numbers of operators of possibly unknown dimension is much more challenging, as is the related problem of distinguishing arbitrary number of states. Any progress made on the general state decision problem has an immediate application on the general operator decision problem. As far as the author is aware, no comprehensive and fast algorithm for choosing optimal measurement directions and calculating the Bayes cost has or  can maybe be found in the discrete general state decision case with arbitrary prior, since the combinatorics of the measurement possibilities and the related posterior calculations increase markedly as the number of states rises and the probe size increases. Only if there is some  conducive structure, is will it be possible to find a  self-contained and efficient solution.

Throughout the present paper we have only considered the cost associated with making
decisions. In any practical situation one must take into consideration
other costs, e.g., the observational cost, state preparation cost, etc., which as constraints might tilt the result in favour of one or another of the strategies. This is  maybe an avenue for more applied research.

As a point of departure from the specific problem considered here, the time spent inside the black box could be made to vary. This would make the situation even more similar to the brachistochrone problem.
 %where the optimal Hamlitonian is determined to change an input state to an orthogonal output state.
%Instead here, as already mentioned, one would determine the optimal input state corresponding to the highest energy eigenvalue difference between the possible Hamiltonians assuming similar eigenvectors or related to the commutator between the different Hamiltonians allowed.

%The result might also shed some light on the  perennial debate about the source of the speed-up in quantum algorithm.
%One candidate is  the $L_2$ norm inherent in quantum probability calculations, another is  entanglement, which as we have shown and is widely known allows one to avoid the law of large numbers effect associated with independent particles.  In our case by combining the action on different probe particles to achieve a linear scaling with $N$ instead of $\sqrt{N}$ in the relevant regime. Naturally, one can ask what effect is more important, and the claim would be that both work together to generate quantum behaviour.

%To avoid this statement sounding to evasive let me add that the first effect is more of a one particle effect and the other associated with multiple particles.
% each having a separate role

 % since the prior is known as well as  the full transition probability, which is all one needs  to calculate the total cost.
Let us recap the logic behind the paper, and explain why it was possible to avoid elaborate calculations. This relies on three insights. First,  the operator decision problem can be mapped into a state decision problem.
 Second, the result of Brody {\it et al.} [5] %\cite{brody} 
 is applicable  that a range of measurement strategies have the same minimal Bayes cost. Third, the close connection with the brachistochrone, which reduces the problem of finding the optimal input state to finding  eigenstates with extremal eigenvalues and combining them in a well-defined way.
%As a caveat, what could reduce the cost further in the entangled case is a strategy, which allows initial measurements to provide feedback for further probe modifications. 

In this paper we always assumed that probe and black box interaction is completed before any measurement takes place.
What happens to the cost function, if   probe modifications are interspersed with measurements will be studied in a separate paper.

%This is why for entangled states the bound calculated is only optimal for the set of allowed strategies.
%As a caveat, what has  not been investigated and could reduce the cost is the case where only part of the entangled state sent through the black box is measured and the result used in some way to modify the rest of the probe sent possibly again through the black box. The authors intuition is that this cannot reduce cost further, but a proof is outstanding.

The author thanks D. C. Brody for stimulating discussions.

\end{document}